\documentclass[aps,pre,two column,floatfix]{revtex4-2}

\usepackage{mathrsfs}
\usepackage{amsfonts}
\usepackage{amssymb}
\usepackage{amsmath}
\usepackage{graphicx}
\usepackage{color}

\newcommand{\be}{\begin{equation}}
\newcommand{\ee}{\end{equation}}
\newcommand{\bqa}{\begin{eqnarray}}
\newcommand{\eqa}{\end{eqnarray}}

\newcommand{\ie}{{\it{i.e.~}}}

\begin{document}

\title{Error Thresholds in Presence of Epistatic Interactions}

\author{D. A. Herrera-Mart\'i}
\affiliation{Universit\'e Grenoble Alpes, CEA List, 38000 Grenoble, France}

\date{\today}

\begin{abstract} 
Models for viral populations with high replication error rates (such as RNA viruses) rely on the quasispecies concept, in which mutational pressure beyond the so-called ``Error Threshold" leads to a loss of essential genetic information and population collapse, an effect known as the ``Error Catastrophe". We explain how crossing this threshold, as a result of increasing  mutation rates, can be understood as a second order phase transition, even in the presence of lethal mutations. In particular, we show that, in fitness landscapes with a single peak, this collapse is equivalent to a ferro-paramagnetic transition, where the back-mutation rate plays the role of the external magnetic field. We then generalize this framework to rugged fitness landscapes, like the ones that arise from epistatic interactions, and provide numerical evidence that there is a transition from a high average fitness regime to a low average fitness one, similarly to single-peaked landscapes. The onset of the transition is heralded by a sudden change in the  susceptibility to variations in the mutation rate. We use insight from Replica Symmetry Breaking mechanisms in spin glasses, in particular by considering the fluctuations of the genotype similarity distribution as the order parameter.
\end{abstract}

\maketitle

\section{Introduction}

Most RNA viruses lack error correction capabilities due to the inherent limitations of RNA-dependent RNA polymerases which, unlike DNA-dependent DNA polymerases, do not have have proofreading abilities \cite{spampinato17} (an important exception to this rule are coronaviruses \cite{smith13, sevajol14}). RNA is less stable than DNA, which also contributes to higher error rates \cite{li99}. Furthermore, DNA repair mechanisms happen mostly in the nucleus, while RNA viruses typically reproduce in the cytoplasm, where there is no access to host's repair enzymes. This lack of error correction in most RNA viruses leads to high mutation rates, contributing to their rapid evolution and adaptability  \cite{duffy08, sanjuan10}. Importantly, the error rate in RNA replication is itself an outcome of selection pressure \cite{duffy18}.

The concept of an ``Error Catastrophe" describes the dynamics of RNA viral populations when mutation rates overwhelm selective pressure, resulting in a loss of genetic information \cite{eigen71}. In this scenario, the viral population transitions into a state where the majority of genotypes have accumulated so many mutations that the genetic information becomes diluted in a pool of defective genomes. The mathematical framework for understanding this phenomenon is provided by the quasispecies model, which is particularly straightforward in the case of single-peaked fitness landscapes, where only one master sequence dominates \cite{eigen89, domingo12}.

Epistatic interactions arise when the effect of one gene is influenced by one or more other genes, and are prevalent in RNA viruses \cite{sanjuan04}. Epistasis can also arise at sub-gene level in RNA viruses \cite{lyons18, lalic12}. This is due to their compact genomes and the fact that protein expression often depends on multiple loci \cite{sanjuan06,elena10}. The presence of epistatic interactions creates complex fitness landscapes characterized by multiple fitness peaks separated by low-fitness valleys, and navigating these rugged landscapes poses a significant challenge for RNA viruses, as beneficial mutations must be combined in specific ways to traverse the fitness valleys and reach new adaptive peaks \cite{colunga23, rokyta11}.

The mechanisms leading to the Error Catastrophe can be understood in the language of continuous phase transitions \cite{leuthausser86,leuthausser87,baake97, bull05}. This analogy helps explain the transition from a state of high fitness to one dominated by defective genotypes. We extend this analogy to rugged fitness landscapes, which are expected to arise whenever the effects of epistasis are strong \cite{kauffman87}. Rugged energy landscapes arise as well in the context of spin glasses, which are magnetic systems with numerous local minima in their landscape arising from conflicting energy requirements \cite{mezard87,nishimori11b}. To study their properties, one can employs the Replica Trick, a mathematical technique that facilitates the analysis of magnetic systems with disordered and conflicting interactions \cite{parisi23}. We apply methods developed in this framework to study epistatic landscapes in RNA viruses, providing a deeper understanding of their evolutionary dynamics and the factors that influence their adaptability, such as the prospect that phenotypes can correspond to classes of genotypes which are separated by fitness valleys. 

The paper is structured as follows. In Section \ref{model} we present statistical-mechanical models of replicator growth to establish the analogy between the Error Catastrophe and ferro-paramagnetic transitions, and we explain how to generalize it to epistatic fitness landscapes. The numerical simulations are presented in Section \ref{results}, where metrics derived from spin glasses are used to diagnose the existence of a threshold. We conclude in Section \ref{discussion} by discussing possible applications and further steps. 

\section{Models for Replicator Growth} \label{model}

The interplay between the mutation rate $\mu$ and the backmutation rate $\mu_{B}$ in RNA viruses is crucial for survival and adaptation. A moderate mutation rate, together with a sufficiently large population size, can foster adaptability and evolution, allowing the virus to thrive in changing environments. Backmutation is the processwhereby a mutated gene returns to its original state. However, the rate of backmutation is typically much lower than the forward mutation rate, meaning that not all harmful mutations can be reversed. Mathematically, this can be understood in terms of graph theory (see Appendix I). Since the coordination number of the mutation graph with genotypes as nodes and mutations as edges  is very large, on the order of the sequence length $L$, mutational paths will diffuse very rapidly and will not go back to the original states. Simple models for the Error Threshold consider a simplified transition matrix of the form in Eq.(\ref{transitionmatrix}), in wich the master sequence is connected to a number of genotypes, which are disconnected among them. This can be argued for on grounds that mutational paths between different genotypes vanish in the sequence length (see Section \ref{results}). However, one drawback of this simplification is that the high-dimensional character of virus difusion in a hypercube cannot be captured \cite{bull05}. Some works have explored workarounds of this problem \cite{elena10, colunga23, takeuchi07}. In order to provide solid grounds for the statistical treatment that follows, we provide in Appendix I analytical and numerical support for reducing a high-dimensional diffusion, potentially in presence of lethal mutations, to the simplified transition matrix in Eq.(\ref{transitionmatrix}).

The size of the viral population also plays a significant role in the relationship betwen $\mu$ and $\mu_{B}$:  in large populations, the impact of harmful mutations can be diluted, as there is a higher chance that some individuals will maintain a lower mutation load and thus higher fitness \cite{manrubia10, tejero10}. Conversely, in small populations, each mutation has a proportionally larger effect on the population's genetic makeup, which can lead to rapid changes in genetic fitness. We define the master sequence molar fraction $\rho$ as the ratio between the number of viruses at the fitness peak and that of all the other genotypes at any point in time. 

\subsection{Landau Theory for the Error Catastrophe}

Whereas it has been suggested that the error catastrophe is a first order phase transition \cite{eigen02}, under the correspondence ``fitness $\Longleftrightarrow$ energy", ``magnetisation $\Longleftrightarrow$ master sequence molar fraction", ``mutation rate $\Longleftrightarrow$  thermal transition probability", and ``backmutation rate $\Longleftrightarrow$ external magnetic field", we show here that one can describe the error catastrophe using the language and machinery of ferro-paramagnetic transitions, which are continuous. The master genotype can be thought as the fully magnetised state (all spins pointing in the same direction, \ie a bitstring made of 0s uniquely), whereas random mutations giving rise to the ``quasispecies cloud" correspond to spin thermal fluctuations. Backmutation, that is, the mutational path that establishes the original sequence, is to be compared with an external magnetic field. It will be useful to define a effective temperature $T$ to compute the transition probability $\mu$ between two genotypes:

\be
\mu(s\rightarrow s') = \exp (- \Delta F / T)
\ee
where $\Delta F = f_s - f_{s'}$ is the fitness difference between two genotypes conected by a transition, potentially by an incorrect nucleoitde replacement. One crucial difference between first and second order phase transitions is that whereas the former rely on a microscopic account of degrees of freedom, the latter form universality classes that obey unversal phenomenological descriptions \cite{nishimori11,nishimori11b}.\\

We now analyse the Error Catastrophe in terms of Landau theory. The fitness matrix:

\be
 {\cal T} = \begin{bmatrix}
f(1- \mu) & f_b\mu_B & \dots & f_b\mu_B \\
f\mu /m & f_b(1-\mu_B) & \dots & 0 \\
\vdots & \vdots & \ddots & \vdots \\
f\mu /m & 0 & \dots & f_b(1-\mu_B)
\end{bmatrix} \label{transitionmatrix}
\ee
 governs the transition probabilities $\vec n_{t+1} =  {\cal T} \vec n_{t}$, where $\vec n = [n_0, n_1, \dots, n_{m}]$. $n_i$ denotes the absolute abundance of genotype $i$, and $n_0$ is the relative abundance of the master sequence, also known as wild-type, which is the genotype with the highest associated fitness. Its normalised dominant right eigenvector is the stationary distribution of the growing population. A canonical obtention of the Error Threshold gives that the critical mutation rate is $\mu_C \approx_{f_b = 1} s$, where $s$ is the fitness advantage over the background landscape, $f= f_b+s$ \cite{bull05}. For a genotype configuration with only one peak at $f=1+s$, we can obtain a critical temperature $T_C  = - \Delta F / \log(1-1/f) \approx -\Delta F / \log(s)$.\\

One important detail, though, is that while the ferro-paramagnetic transitions are static, in the sense that the state is fully determined by the parameters and stationary in time, the mathematical framework which describes Error Catastrophe entails a population variability which depends on the fitness of the dominant eigenvectors. If the fitness associated to an eigenvector of the population matrix is larger (smaller) than one, this eigenvector will grow (decrease) with time. Therefore, this analogy holds only away from the mutational meltdown regime, that is, all considered states need to have a fitness equal or larger than 1, so that populations are maintained and molar fractions are stabilised in time \cite{bull05}(see Appendix I).\\

\begin{figure}[b!]
  \includegraphics[width=\linewidth]{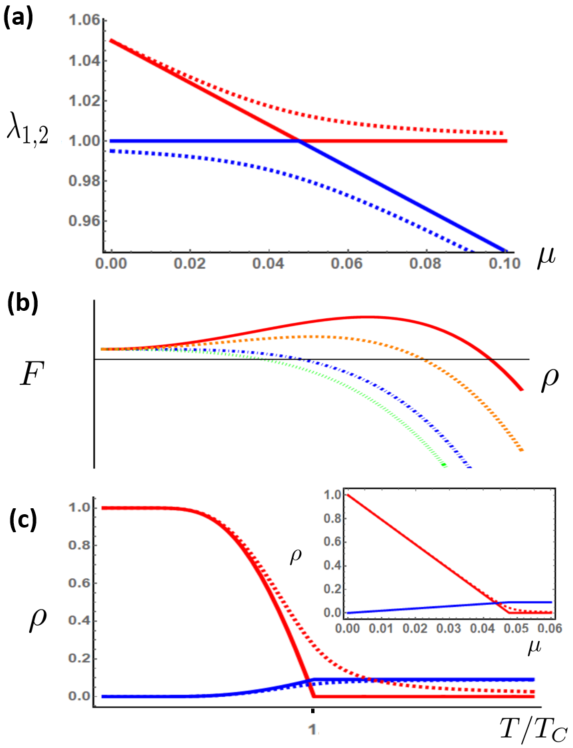}
  \caption{The Error Catastrophe as a second order (continuous) phase transition. Using a simple model with a single peak of fitness $f = f_b + s$ against a background of baseline fitness $f_b$. Here $f_b = 1$, $s = 0.05$ and $\mu_B = 0$ (solid lines) or $\mu_B = 5 \times 10^{-3}$ (dashed lines). {\bf (a)} The two dominant eigenvalues show an anticrossing for non-zero backmutation rates $\mu_B$. In the strong selection, weak mutation regime, the fitness decreases linearly with $\mu$, whereas for large $\mu$ the dominant eigenstate is the uniform distribution of all genotypes (excluding the master sequence), which is the stationary distribution in the weak selection, strong mutation regime.{\bf(b)} Landau functional for the fitness. At low temperature (mutation rates), $\alpha<0$ and the maximum of the functional happens at a non-zero value for the master sequence molar fraction. For high mutation rates, the fitness is maximised for $\rho = 0$. Here $\alpha/\beta \in (0.5, 0, -1, -2)$ for dotted green, dot-dashed blue, dashed orange dotted and solid red, respectively.{\bf (c)} The order parameter as a function of the effective temperature ($T_C  \approx -\Delta F / \log(s)$). Different values of $\mu_B$ contribute to smoothing out the transition. Inset: the more familiar stationary distribution as a function of $\mu$. Solid line corresponds to $\rho = n_0/\sum n_i$, the dashed line corresponds to the average of $n_i/\sum_j n_j, i>0$.}
  \label{fig1}
\end{figure}

Diagonalising the transition matrix for the dominant eigenvalues in the limits $\mu \ll \mu_C$ and $\mu \gg \mu_C$ correctly reproduces the asymptotic states of ``strong selection, weak mutation" (completely magnetised) and ``weak selection, strong mutation regime" (completely demagnetised state), as shown in Fig.\ref{fig1}{\bf(a)}. This constitutes a strong hint that the average fitness above the baseline $\langle F \rangle = \sum_s f_s p_s$ of a viral population near the error catastrophe can be expressed as a power series expansion in terms of an order parameter $\rho = p_0$, with $p_s = n_s / \sum_j n_j$. For a single-peaked fitness landscape with fitness $f_b$ everywhere except of at the peak, the fitness functional  $F(\rho, T)$ can be written as:

\be
F(\rho, T) = f_b - \frac{\alpha(T)}{2}\rho^2 - \frac{\beta}{4}\rho^4 + \cdots
\ee
where $\alpha(T)$ is a temperature-dependent coefficient. The coefficient $\beta$ is  positive to ensure stability of the fitness around its maximum \cite{nishimori11}. Here the fitness is maximised, in contrast with the free energy in a ferro-paramagnetic transition, which is minimised by the magnetisation order parameter (see Fig.\ref{fig1}{\bf(b)}). The coefficient  $\alpha$ changes sign at the critical mutation rate $\mu_C$:
$$
\alpha(T) = \alpha_0 (\mu - \mu_C) = a_0 (T - T_C)
$$
where $\alpha_0$ and $a_0$ are positive constants (since $\mu$ is a monotonic function of $T$). At mutation pressures above $\mu_C$ (\ie, temperatures above $T_C$),  $\alpha > 0$ becomes positive and the genetic information of the master sequence gets diluted in the total population,  $\rho \rightarrow 0$. Below  $\mu_C$ ($T < T_C$),  $\alpha < 0$ and the virus population is clustered around the master sequence, \ie $\rho > 1$ (see Fig.\ref{fig1}{\bf(b,c)}).

The ``mutational susceptibility" $\chi$ measures the response of the master sequence molar fraction to an fluctuating backmutation rate $\mu_{B}$. It is given by:

\be
\chi = \left. \frac{\partial \rho}{\partial \mu_{B}} \right\vert_{\mu_{B}\rightarrow 0}
\ee
In this case, the ``external parameter" is the backmutation rate $\mu_B$.  To obtain $\chi$ in Landau theory, we add a term $\rho\mu_{B}$  to the fitness functional functional, since the backmutation rate $\mu_{B}$ can be thought as being thermodynamically conjugate to the order parameter $\rho$:
$$
 F(\rho, T, \mu_B) = F(\rho, T)  + \rho\mu_{B}
$$

Slightly above $\mu_C$, in the weak selection, strong mutation regime (\ie paramagnetic phase), it is possible to approximate $\rho$ for small $\mu_{B}$ (see Appendix I):
$$
 \rho \approx \frac{\mu_{B}}{\alpha}
$$

\begin{figure}[ht!]
  \includegraphics[width=\linewidth]{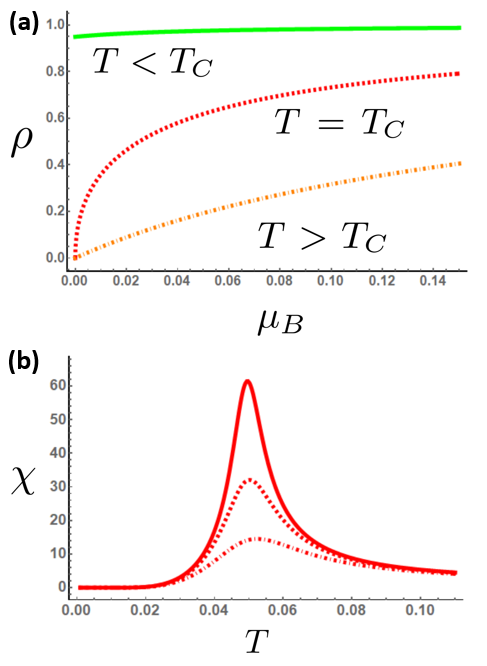}
  \caption{Critical behavior of the single peaked landscape at the mutation threshold. {\bf(a)} The order parameter $\rho$ dependence on backmutation rate $\mu_B$. The three regimes of interest: strong selection regime ($T < T_C$, green solid line), at the critical temperature ($T = T_C$, red dot-dashed line) and strong mutation regime ($T > T_C$, dashed orange line) exhibit a behavior analogous to that of a continuous ferro-paramagnetic transition. In this case, the first order phase transition is absent because $\rho$ is defined positive, contrary to magnetization, which can change signs. {\bf(b)} The ``mutational susceptibility" $\chi$ diverges at the error treshold. This can be understood in the following manner: deep inside the strong selection-weak mutation regime, backmutation does not play a significant role because the viral population is already concentrated around the fitness peak. On the other hand, in the weak selection-strong mutation regime, the mutation rate overwhelms any change in the backmutation rate. At the critical mutation rate, small fluctuations of $\mu_B$ have an enormous effect on the order parameter. $\mu_B \in (10^{-2}, 5 \times 10^{-3},10^{-3})$}
  \label{fig2}
\end{figure}

Since  $\alpha =  a_0 (T - T_C)$:

\be
 \chi =  \left. \frac{\partial \rho}{\partial \mu_{B}} \right\vert_{\mu_{B}\rightarrow 0}  = \frac{1}{a_0 (T - T_C)}
\ee
which shows that the susceptibility diverges as $T$ approaches $T_C$, following a Curie-Weiss law, as shown in Fig.\ref{fig2}{\bf(a,b)}.

\subsection{Simplified Model for Epistatic Interactions}

Ever since the inception of population genetics it has been emphasized that epistatic interactions, that is, effects of combinations of genes rather than just genes' individual effects, is the determinant factor in the study of fitness landscapes \cite{wright31}.

The underlying physical mechanisms by which sets of genes tend to mute together is a still debated subject, and it depends on the kind of replication considered. For instance, during meiosis in the diploid case, DNA condensation and pairing due to electrostatic interactions, or possibly protein-mediated interactions \cite{cherstvy05, cherstvy13}, seem to help understand why distant genes in the DNA sequence mute together. However, the concept of replicator at genotype level is difficult to argue for in diploid, sexed reproduction, given allele dominance and randomisation of the genetic code into several gametes during meiosis. In the haploid case, purely stochastic models of gene (or even intragene) epistasis are particularly simpler: while additive models give that the variation in fitness of a mutation $s\rightarrow s'$ at $K$ nucleotides comes given by $\Delta F = \sum^K_j \delta f_j$, where $\delta f_j$ stand for fitness variations per nucleotide, epistasis require that interactions between nucleotides (or sets thereof) be taken into the account in the computation of $\Delta F$. In particular, the \emph{NK model} \cite{kauffman87} allows to quantitatively tune the effects of epistasis in a way that naturally leads to fitness peaks separated by valleys, and is used to study the complexity of fitness landscapes and the role of epistatic interactions in biological systems. Its main parameters are $N$ (number of loci in the genome) and $K$ (degree of epistasis). The fitness of a particular genotype $s$ is determined by the contributions from each locus and its interactions with $K$ other loci: 

\be
F_s=\frac{1}{N}\sum^N_i f_i(l_{i0},l_{i1},l_{i2},…,l_{iK}),
\ee
where $f_i$ is the fitness contribution of locus $i$, which depends on the state of locus $i$ and the states of $K$ other loci $(l_{i0},l_{i1},l_{i2},…,l_{iK})$. There are two limit cases: when $K=0$, each locus contributes independently to the fitness, resulting in a smooth additive fitness landscape. As $K$ increases to a maximum $K = N-1$, the fitness landscape becomes extremely rugged with many local optima. Interestingly, the backmutation rate is influenced by the epistatic interactions.

\begin{figure}[ht!]
  \includegraphics[width=\linewidth]{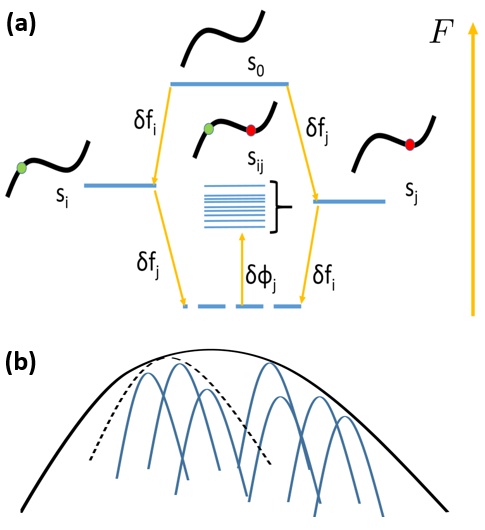}
  \caption{Pictorial representation of the simplified NK model {\bf(a)} Antagonistic epistasis, which is strictly positive for deleterious mutations, results in landscapes where concomitant mutations at different loci can result in genotypes with higher fitness than expected from an additive model. These epistatic genotypes can have a fitness that even surpasses that of the genotypes resulting from single mutations, separatedly. In this model, $\mu$ depends on the ratio between the effective temperature and the fitness barrier, $h_i / T$. Higher fitness differences will result in smaller mutation probabilities, which means that at low temperatures, mutations will tend to cluster in the loci that increment fitness, \ie those affected by epistatic interactions. At high temperatures, mutations will be evenly distributed along the whole sequence. {\bf(b)} Representation of a rugged landscape in which the genotypes at the peak of the fitness maxima that lie close together can be considered as belonging to the same phenotype (thick dashed line). For high enough temperatures, the whole configuration space is sampled (thick continous line).}
  \label{fig3}
\end{figure}

One drawback of the NK model, when it comes to simulatability, is its intrinsic randomness. That is, given $N$ and $K$, the loci and interaction strenghths are distributed at random, which makes it challenging to identify at the outset where the fitness maxima lie, since interactions are sampled at random. In order to fully characterise these maxima, an exhaustive search or computationally expensive Monte Carlo sampling must be carried out, which severely limits the size of systems ameable to full analysis. We circumvent this limitation by reducing the expressive ability of the NK model and allowing only a reduced set of epistatic interactions. We only consider fitness variations of the sort:

\be
\Delta F = \sum^K_i \delta f_i + \sum^{n}_j \delta \phi_j
\ee
where $ \delta f_i < 0, \forall i$ stem from deleterious non-lethal mutations and $\delta\phi_j = \sum^{K/n}_i \epsilon^{(j)}_i > 0$ represents the positive, epistatic interaction between nucleotides (see  Fig.\ref{fig3}{\bf(a)}).

It has been observed that epistasis in RNA viruses is overwhelmingly of antagonistic nature \cite{lalic12}, that is, given the presence of one or several deleterious mutations, new epistatic mutations are expected to be \emph{less} harmful to fitness than they would be in the absence of the previous mutations. If antagonistic epistasis is strong, the likelihood of backmutations is decreased because the genotype mutant is separated from the original one by fitness valleys. Genetic recombination, which is prevalent in positive strand viruses, is another mechanism leading multiple peaked phenotypes \cite{barr16}. Although recombination is a separate mechanism from mutation, it can be accounted for in this model by increasing the ratio $K/n$, which is a measure of mutation clustering. Antagonistic epistatic behavior can be reproduced by ensuring that mutations will tend to accumulate at loci that maximise fitness, or decrease it minimally. The main motivation for adopting this simplified model is that it allows for direct sampling of fitness maxima, which are fixed beforehand, as opposed to the canonical NK model, where due to the random interactions it is necessary to sweep configuration space to completely characterise the minima. Throughout this work we will neglect the effects of the interplay between epistasis and mutation rates  and assume that the former are tunable parameters.\\

Mathematically, we impose that the mutation rate at locus $i$ is $\mu_i = \exp(-h_i / T)$, where the finess barrier $h_i$ is small at loci subject to positive epistatic interactions and large everywhere else. This allows to establish beforehand the location of fitness maxima, which is necessary to carry out Monte Carlo simulations for large $L$, as explained in next section.

The $K$ mutations are distributed evenly in $n$ loci, each hosting $K/n$ mutations. The value $K$ determines how many different mutational paths can connect two different fitness peaks. For a given $K$, the number of mutational paths from the zero mutation state is $K !$. The number of paths increases factorially, but each path has a probability of $\sim p_\textrm{no LM} \times \mu^K(1-\mu)^{L-K}$, where $p_\textrm{no LM} = (1 - p_\textrm{LM})^K $ is the probability of the path encountering zero lethal mutations.  So the probability of each path vanishes exponentially for small $K/L$. The fitness peaks are thus effectively disconnected, which is the reason behind the loss of ergodicity. \\

Notice the resemblance with spin glasses, in which there are local terms (like the $\delta f_i$s) and interaction terms ($\delta \phi_j$). The degree of epistasis $K$ in the NK model determines how rugged the fitness landscape is, similar to how the interaction strength and randomness determine the complexity of the energy landscape in spin glasses. In spin glasses the total magnetisation is not a well-defined order parameter. In the same spirit, we will not be using  $\rho$ defined in previous subsection to study the multiple peak regime, since it will itself become a random variable depending on the location of the fitness peaks.  Instead, we will used an order parameter based on genotype overlaps, inspired on the spin glass susceptibility for disordered magnetic systems.\\

\section{Numerical Simulations and Results} \label{results}

In spin glasses, Replica Symmetry refers to the theoretical assumption that all instantiations of the system with random couplings sampled from a distribution (\ie system replicas), which are necessary to estimate the free energy, are equivalent \cite{mezard87,nishimori11b,parisi23}. This symmetry implies that the system is in a regime where a single energy minimum dominates the phase space.  This can arise at either higher temperature, where the system's behavior is typically less complex, or in energy landscapes where a single energy minimum dominates. In spin glasses, as the temperature decreases, the system may transition into a phase where the Replica Symmetry assumption no longer holds, because instead of a single (free) enery minimum, many energy minima have similar thermodynamic properties and are macroscopically equivalent. However, these minima are separated by energy barriers, leading to a rugged energy landscape which leads to a breakdown of ergodicity, meaning that the system gets effectively trapped forever (over observational timescales) in one energy minimum. When this happens, each replica will typically end up in a different energy minimum, \ie Replica Symmetry Breaking (RSB) occurs.

\subsection{Replica Symmetry Breaking in Viral Populations}

Under the aforementioned ``fitness $\Longleftrightarrow$ energy" correspondence, RSB can be observed in viral populations. This concept is not to be conflated with the quasispecies description of a viral population, in which genotypes fluctuate within same phenotype. In the case of RSB, the existence of several fitness peaks indicates a genotype diversity which itself can lead to different phenotypes (distant local maxima). Define the genotype overlap as:

\be
q(s,s') = \frac{1}{L}\sum_j \delta_{s_j, s'_j} 
\ee
which takes values between $0$ and $1$. However, empirically we will only consider values of $q$ above $1/2. $This is because, unlike spin models which are symmetric under the label swapping ($1 \longleftrightarrow -1$), the coexistence of genomic and antigenomic sequences, which have the same genetic information but use complementary nucleotides and correspond to the symmetry ($1 \equiv 0$) halves the size of the configuration space.

\begin{figure}[ht!]
  \includegraphics[width=\linewidth]{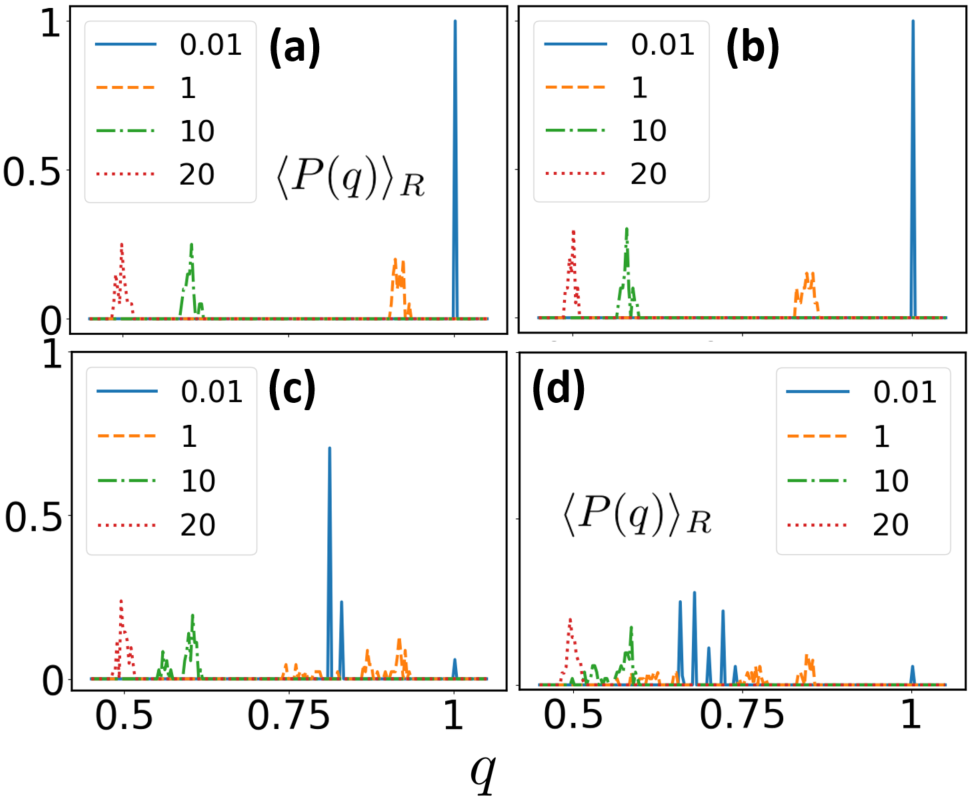}
  \caption{Overlap distributions $\langle P(q) \rangle_R$ averaged over 100 different instantiations of the simplied NK model, with 30 fitness maxima each, for sequences of length $L = 5000$. A departure from a delta peak signals the onset of Replica Symmetry Breaking.  Increasing ratios of temperature over the barrier $T/h$ are shown in the legend. As the effective temperature increases, the overlaps tend to shift towards the left of the x-axis: they transition toward a fully disordered state.  {\bf(a)(b)} The cutoffs $q_\textrm{cutoff} = 0.95$ impose a high lower bound on the similarity measure between genotypes. {\bf(c)(d)} The cutoffs $q_\textrm{cutoff} = 0.75$ denote an initially spread out state over many minima (as can be seen from the multiple peaks of the overlap distribution at $T/h = 0.01$. {\bf(a)(c)} Percentage of epistatic loci: $25\%$ {\bf(b)(d)} Percentage of epistatic loci: $50\%$ }
  \label{fig4}
\end{figure}

\begin{figure}[ht!]
  \includegraphics[width=\linewidth]{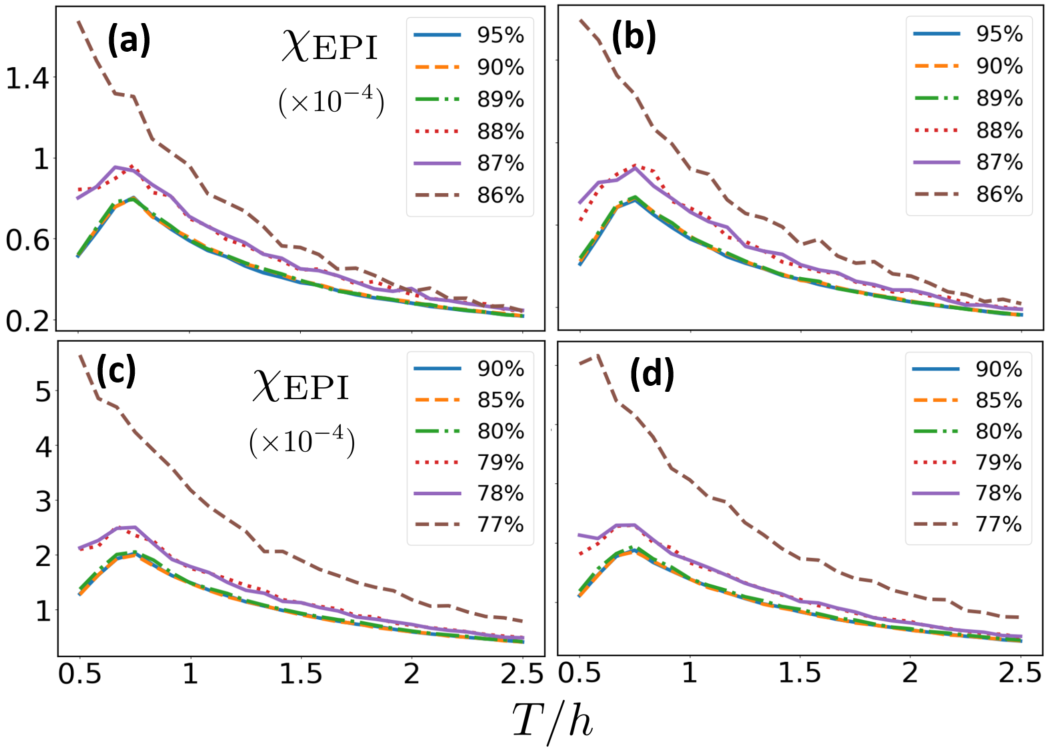}
  \caption{Epistatic susceptibilitiy $\chi_\textrm{EPI}$ as a function of the ratio betwen the effective temperature and barrier $h$ for epistatic loci, for different cutoffs. It is assumed that the barrier at non-epistatic loci is 10 times larger, \ie $h_\textrm{non-EPI} =10h$. Each point was obtained from running a simulation with 30 fitness maxima and 1000 instantiations. {\bf(a)(b)} Percentage of epistatic loci: $25\%$ {\bf(c)(d)}Percentage of epistatic loci: $50\%$.  {\bf(a)(c)} $L=5000$ {\bf(b)(d)}$L=10000$. The cusp in the susceptibilities reveal the existence of a transition from a strong selection low mutation regime (in the presence of epistatic interactions) to a strong mutation low selection regime. For a fixed percentage of epistasis, as the cutoff increases (as shown in percentages in the legend), a critical temperature becomes visible in the diagram. Above this temperature, the population is in a strong mutation / disordered state.}
  \label{fig5}
\end{figure}

The signature of RSM is the spreading of the overlap distribution $P(q) = \sum_{s,s'} \delta(q -  q(s,s') ) $ and the departure from a delta function \cite{parisi23}(see Appendix II and Fig.\ref{fig4}). Likewise, a genotype overlap distribution that deviates form a highly concentrated distribution indicates that the system has entered a phase into which several phenotypes coexist in the low mutation rate regime (see Figs.\ref{fig4}-\ref{fig5}). The overlap distribution $P(q)$ is itself averaged over several instantiations (replicas) of the simplified NK model to give $\bar{P(q)} = \langle P(q) \rangle_R$.

The Monte Carlo sampling performed to estimate the distributions consisted on a simple Metropolis-Hasting algorithm with in which we assume that the fitness barrier at non-epistatic loci is 10 times larger than for epistatic loci, \ie $h_\textrm{non-EPI} =10h$. This means that for low effective temperatures (strong selection, weak mutation regime), the algorithm only samples genotypes which are of antagonistic epistatic nature, \ie genotypes with typically higher fitness. This corresponds to sampling around a local optimum. In the high temperature regime, the genotypes of all fitnesses are sampled.

In this extension of the strong selection, weak mutation regime, a new way of testing the response of the viral population to changes in its environment is necessary. Very much like in the case of the spin case, where the magnetic susceptibility is generalized to a spin-glass susceptibility,  the ``mutational susceptibility", which measures the impact of variations of the backmutation rate on the master sequence molar fraction, must be generalized into an ``epistatic susceptibility", which measures how the fitness peak overlaps change as a function of fluctuations in the epistatic interactions. This ``epistatic susceptibility" is computed as the replica average ($\langle .\rangle_{\bar P}$) of the variance of overlap distribution:

\be
\chi_{EPI}(T) = \frac{1}{T}\left( \langle q^2 \rangle_{\bar P} - \langle q \rangle^2_{\bar P} \right)
\ee

Similar quantities have been defined for glassy transitions in RNA molecules, albeit in different contexts \cite{pagnani00, andreanov10}. If a particular sequence at a fitness peak is being subject to small mutation pressure, the mutated genotypes will tend to cluster within a maximum Hamming distance of the fitness master sequence and will form a ``cloud" around it. Intuitively, it will form a local version of the single-peaked landscape, in which small changes in the backmutation rate can substantially contribute to fitness variations. Note that several fitness peaks could conceivably correspond to the same phenotype, especially if the peak sequences are very similar (which would correspond to epistatic interactions arising within the mutant cloud).  Estimates for the percentage of loci that are subject to epistatic interactions ranges from $40\%$ to $80\%$ \cite{lalic12, rokyta11}. Throughout this work, we assume that a conservative $25 - 50\%$ of the genotype to sustain epistatic mutations.\\

Beyond a mutation threshold (see Fig.\ref{fig5}), mutations will accumulate evenly at all non-lethal loci and the viral population becomes insensitive to the fitness and epistasis of individual genotypes. It is, so to speak, as if the high effective temperature smoothes out the fitness landscape. As expected, the behavior of $\chi_{EPI}(T) $ closely resemples that of the single-peak landscape. 

At low temperatures, the mutations in our model happen almost exclusively at loci subject to epistatic interactions, since in the presence of antagonistic epistasis, mutations at epistatic loci are likely to lead to fitter sequences than mutations at random loci. In this case, two different possible behaviors arise, depending on the balance between the percentage of epistatic loci in the genotype and the strength of the fluctuations across different replica.

Imposing a cutoff $q_\textrm{cutoff}$ on the initial overlap ditribution sets a upper bound on the fluctuations of the distribution $\bar{P(q)}$. For high cutoffs,\ie $q_\textrm{cutoff} \lessapprox  1$, the ``mutational susceptibility" is computed over a reduced number of nearby peaks (see Figs.\ref{fig5} and \ref{fig3}{\bf(b)}). This susceptibility qualitatively resembles the one that is obtained in the single-peak landscape, albeit with a finite cusp, and can be interpreted as quantifying the effect of mutations in one or several peaks that correspond to a single phenotype. This is due to the fact that the overlap variance remains small, since only similar genotypes are being considered (see Appendix III). Conversely, if the virus population is allowed to explore larger swathes of configuration space, as could arise for instance in the case where the mutation pressure decreases progressively, it will retain the ability to choose the fitness peak that is more suitable for a given mutation and backmutation rates ($q_\textrm{cutoff} < 1$). In this case the ``mutational susceptibility" is computed by averaging over arbitrarily different peak sequences. As a result, the variance of the overlaps remains well above zero, and $\chi_{EPI}(T) \propto \frac{1}{T}$ diverges at low temperatures, as shown in Fig.\ref{fig5}.

This closely resembles the magnetic case, where a change of slope in the susceptibility is the hallmark of a phase transition \cite{nishimori11b}: whereas above the critical mutation rate the population is a ``weak selection, strong mutation regime", which is analogous to a paramagnetic phase in spin glasses, at low mutation rates the population is in a ``strong epistatic interaction regime", analogous to a spin ice, in which short-range order is maintained. We numerically calculated the ``epistatic susceptibility" by computing the genotype overlap distribution averaged over several replicas. For high enough overlap cutoffs, one clearly sees the emergence of a transition, \ie the emergence of a cusp in $\chi_{EPI}(T)$ (see Fig.\ref{fig5}). This change of slope serves the purpose of heralding a phase transition, which can be understood as a generalisation of the error catastrophe for epistatic landscapes.

By restricting the amount of minima that are considered in the calculations of the susceptibility, and in particular by considering only minima that are close in Hamming distance, one is effectively reducing the strength of epistatic interactions and the zooming into the ``strong selection, weak mutation regime" resembles more that of a single-peaked landscape \cite{parisi23}. This is confirmed by the progressive concentration of the genome overlaps around a delta function. The transition from one susceptibility to the other can be implemented by progressively imposing a Hamming distance cutoff between fitness peaks (see  Fig.\ref{fig5}).

\begin{figure}[ht!]
  \includegraphics[width=\linewidth]{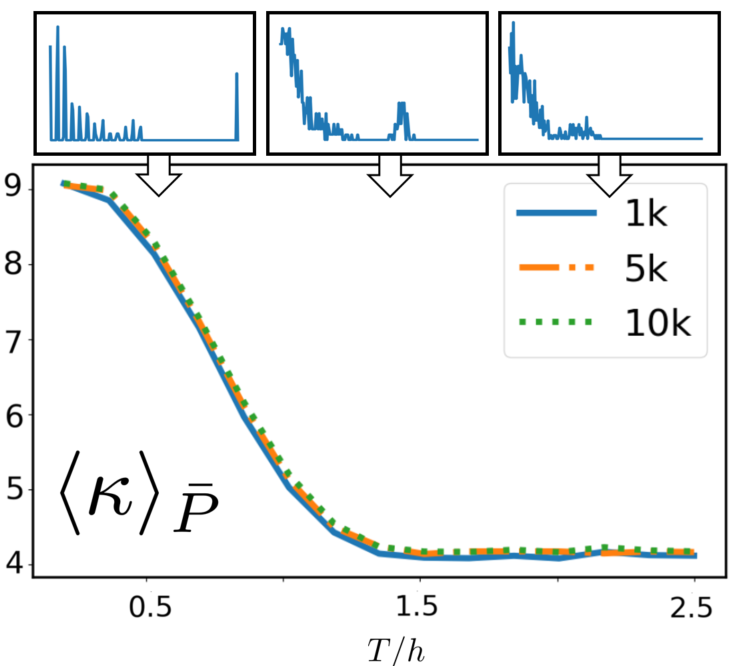}
  \caption{Replica-averaged kurtosis of the overlap distributions, $\langle \kappa \rangle_{\bar P}$. Percentage of epistatic loci:$50\%$, cutoff $ q_\textrm{cutoff} = 0.5$. At low mutation rates, we observe a leptokurtic behaviour, which is consistent with the tradeoff between the bulk of overlaps (toward the value of $q=1/2$) from states in different maxima (a delta-like peak at large values of $q$). For high enough mutation rates (temperature beyond the threshold), the distributions appear to be more mesokurtic. This is expected since mutations cause the states in local maxima to become relatively randomized, which pulls the remnant of the delta function (rightmost-peak) toward the left, as shown in the insets. Insets: typical overlap distributions in three different regimes ${\bar P (q)}$, in the range $q \in (0.5,1)$. Left: strong selection in the presence of epistasis. Right: strong mutation. Center: intermediate regime.}
  \label{fig6}
\end{figure}

In order to provide further evidence for the existence of a phase transition, we performed a simple finite-size scaling analysis of the kurtosis (using Fisher's definition) of the overlap distribution. This kind of analysis resembles that of the Binder cumulant \cite{binder81} in lattice systems. We considered sequences of increasing length, $L \in (1k,5k,10k)$ (See Fig.\ref{fig6}), and found that the overlaps transition from a regime in which the genotypes tend to be clustered (strong selection, weak mutation) to one in which the overlaps are relatively spread out (weak selection, strong mutation).

\section{Discussion and Outlook} \label{discussion}

We have provided a framework that allows to generalize the analysis of the Error Catastrophe to rugged energy landscapes.  The analogy with continuous phase transitions in magnetic systems has proved to be a fruitful one in the analysis of epistasis and error thresholds for mutability in quasispecies models.\\

We have shown that there exists a threshold above which the virus population remains in a strong mutation regime in epistatic landscapes. Below this threshold, we have generalized the behavior from a single-peak landscape (additive mutations) to a multiple peaked landscape, which is arguably a more complex and realistic scenario for virus evolution. We have therefore contributed to enlarging the applicability of the theory underlying the Error Catastrophe to more realistic cases (assuming large viral populations away from the mutational meltdown threshold). \\

Backreaction rates are dependent on a host of diverse factors, such the prevalence of lethal mutations, the mutation pressure itself or the sequence configuration associated to a fitness peak. We have assumed that they are an independent parameter, as can be done in spin systems, which is debatable. However, this independence, a figment as it may appear, can be useful to reason about drug safety and emergence of mutant strains. Mathematical models for quantifying safety of treatments based on mutagentic drugs asume a dynamics which is then fitted to experimental parameters \cite{hassine22,lobinska23}. This may or may not be representative of real effects. The calculation of overlaps at any given point in time allows to monitor the effect of mutagenic drugs on control populations in more realistic, model-free scenarios. For instance, this approach could provide a way to screen epidemics in the absence of abundant ressources, as can be the case during the rising curve of an outbreak. If one assumes that it is possible to perform pooled genomic analysis, then computing the overlap matrix between sequenced viruses provides valuable, real-time snapshot of the viral population, emergence of strains, and responses to treatments or changing external factors.\\

We touched repeatedly on the concept of diffusion of a population of replicators in a high-dimensional hypercube. While this is not new, the application of spectral graph theory to determine that an Error Threshold can exist in the presence of lethal mutations is a novel result, to the best of our knowledge. An interesting avenue of research would be to characterize whether and how different kinds of fitness landscapes (\ie additive, multiplicative, rugged...) lead to a threshold. Another possible improvement of this work would be to consider epistatic models with complexity higher than the simplified NK model presented here. This would potentially lead to a complete calculation of Binder cumulants, albeit at the expense of heavier Monte Carlo sampling. We leave these two questions for further work.

\section{Acknowledgements}

I would like to thank Stefano Mossa and Rob Whitney for technical remarks. Early motivation for this project came up during discussions at the East African Institute for Fundamental Research.

\bibliographystyle{unsrt}

\newpage
\widetext

\section*{Appendix I: Existence of an Error Threshold in the Presence of Lethal Mutagenesis}

It has been argued that lethal mutagenesis is incompatible with the existence of an error threshold in viral populations. The underlying reason for this argument is that lethal mutagenesis necessarily implies a reduction of the viral population since lethal mutants, which dominate over non-lethal ones, do not reproduce and act as a population sink \cite{wilke05}. The mathematical model for the error catastrophe needs a finite non-vanishing ratio between the fit and deleterious genotypes. \\

While this argument might hold for high percentages of lethal mutations, we show that there is a regime in which this need not be the case generally. If at each time step there genotyope has a non-zero probability of transitioning into a deleterious non-lethal mutation with fitness above 1, this will sustain a ``escape route" that will circumvent the lethal genotypes (corresponding to absent vertices in the hypercube graph). In other words, the important factor is not whether the amount of mutations that are lethal at each time step is dominant, but rather wethere there are \emph{at least} a few mutations which are viable. If this is the case, the virus can continue to evolve, albeit less efficiently.\\

In order to see how this might happen, it is important to consider the full-dimensional model that leads to random diffusion in a hypercube: a complete mathematical description of the Error Catastrope in viruses of genomic length $L$ involves considering every sequence in $2^L$ configurations. A simple model which considers only single mutations between genotypes comes given by the the adjacency matrix of the hypercube graph weighted with the fitness of each genotype.

$$
{\cal T} _\textrm{hypercube} = \left( \textrm{diag}([1-\mu_{0},1-\mu_{1}\dots 1-\mu_{2^L}]) + \frac{1}{L} A_\textrm{hypercube} \textrm{diag}([ \mu_{0}, \mu_{1}\dots \mu_{2^L}]) \right)  \textrm{diag}([f_{0},f_{1}\dots f_{2^L}]) 
$$
where  $A_\textrm{hypercube}$ is the hypercube graph adjacency matrix and $\textrm{diag}([...])$ denotes the diagonal matrix. \\

Spectral graph theory provides valuable tools to analyse graph connectivity in terms of the eigenvectors and eigenvalues of the Laplacian matrix. We compute the spectral gap of the Laplacian matrix $\mathcal{L}$ of the $L$-dimensional hypercube graphs subject to vertex percolation (that is, vertices are absent with probability $p$), as  a way to model lethal mutations. The Laplacian Matrix is defined as $\mathcal{L} = D - A$, where $D$ is a diagonal matrix encoding the coordination number of each vertex. We show numerically that the $L$-dimensional hypercube remains connected, that is, there is a path connecting the "000...0" string with "111...1" string (or local variations thereof), provided that the lethal mutation rate is below a threshold. The spectral gap (the difference between the first two eigenvalues) is an upper bound of the Cheeger constant $h_C$, which is strictly positive ($h_C > 0$) if the graph is connected, and zero if it consists of disconnected subgraphs. Defining $\sigma_{12} = \lambda_2 - \lambda_1$ as the spectral gap of the $L$-dimensional hypercube graph, we use the well-known relation:
$$
2 h_C \geq \sigma_{12} \geq \frac{h^2_C}{2 L}  
$$
to infer that the hypercube remains connected even for significant abunance of lethal genotypes. Small values of $h_C$ imply that there is a graph bottleneck, which means that two subgraphs are connected by a small number of edges. \\

We simulated virus diffusion in different landscapes (with a single fitness peak) for strings of lenths up to 15, and found good agreement between the behavior of the spectral gap (as a function of vertex percolation probability) and the long term molar fraction of the dominant peak sequence. The transition matrix of the hypercube can be written as ${\cal T} _\textrm{hypercube} = [I - \mu(I - \frac{1}{L}A)] diag([f_0,\dots f_{2^L}]) =  [I - \mu(\mathcal{L}/L)] diag([f_0,\dots f_{2^L}])$, meaning that they are not diagonal in the same basis.\\

\begin{figure}[ht!]
  \includegraphics[width=\linewidth]{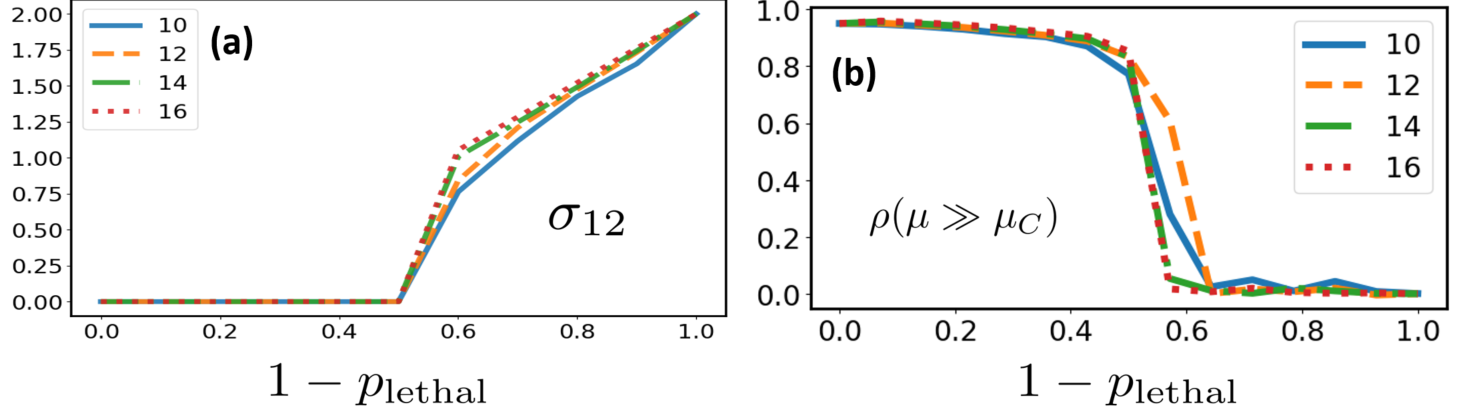}
  \caption{ Simulation of $L$-dimensional hypercube, which models mutations in sequence space of legth $L=10$ up to $L=16$, for different percolation probabilities $p_ \textrm{lethal}$, which corresponds to the prevalence of lethal genotypes. Percolation removes a vertex from the graph. Equivalently, it reduces the fitness to zero and removes all links \emph{from} the lethal genotype. {\bf(a)} The spectral gap $\sigma_{12}$ of the Laplacian matrix is a measure of graph connectivity. A non-zero value indicates that the graph is connected. Small values indicate that there is a connectivity bottleneck between subsets of vertices. {\bf(b)} Master sequence molar fraction for $\mu\gg \mu_C$ for different lethal mutation prevalences. As expected, the molar fraction remains near 1 for disconnected graphs, and then it abruptly decreases around the same value in which the spectral gap indicates that the hypercube is connected.}
  \label{figI2}
\end{figure}

It is possible to obtain information from the asymptotic behavior of $\rho = n_0 / \sum_j n_j$ for large mutation rates. If lethal mutations dominate, then $\rho$ is expected to remain constant since the fitness peak is disconnected from most of the other genotypes. Conversely, if $\rho$ decays at some value of $\mu$ (which will typically be larger than the original critical mutation rate), then one can infer that the master sequence has been diluted in the total population since the graph remains connected. This behavior is observed in  Fig.\ref{figI2}{\bf(b)}. Importantly, the value of $\mu$ for which the collapse of $\rho$ is observed will typically be lower than the mutational meltdown point, \ie the corresponding eigenvalue will be larger than 1 (the effective mutation rate can increase if transitions between genotypes with Hamming distance above 1 are allowed).

\begin{figure}[ht!]
  \includegraphics[width=\linewidth]{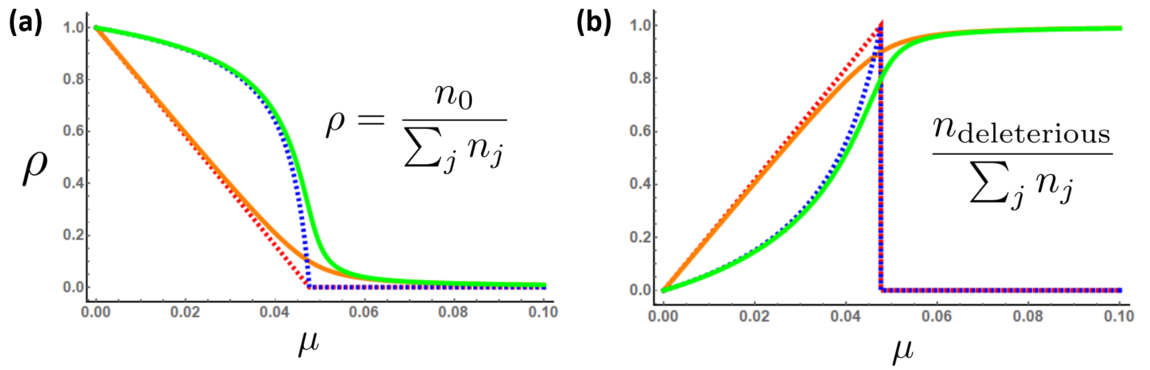}
  \caption{Threshold in the presence of lethal mutations. For a non-zero backmutation rate $\mu_B$, the population of the lower fitness genotypes tends to 1, even when lethal mutations are widely prevalent. {\bf(a)} Population of master sequence for $r=0$ ($r=0.8$) plotted in red (blue) dashed lines. In the presence of the backmutation, the master sequence population is plotted in solid orange (green) {\bf(b)} Population of low fitness non-lethal genotype for $r=0$ ($r=0.8$) plotted in red (blue) dashed lines. In the presence of the backmutation, the population for these deleterious is plotted in solid orange (green). In all cases $\mu_B = 10^{-3}$}
\label{figI3}
\end{figure}

The simplest model for the error catastrophe in the presence of lethal mutations can be built by grouping the microscopic into three collective states: the master sequence, the non-lethal deleterious mutations and the lethal mutations :
$$
M = \begin{bmatrix}
f (1-\mu ) & \frac{\mu_B}{2} & 0 \\
f \mu  (1-r) & 1-\mu_B & 0 \\
f \mu r & \frac{\mu_B}{2} & f_L \\
\end{bmatrix}
$$
where $r$ is a proxy for the ratio of lethal mutations, which all have fitness strictly lower than 1, $f_L < 1$.  This coarsegraining allows a treatment in terms of effective mutation rates. It can be seen that the existence of a population sink stemming from the presence of lethal mutations amounts to a renormalisation of the non-dominant fitness and the backmutation rate (see  Fig.\ref{figI3}). For high percentages of lethal genotypes (as $r \rightarrow 1$), the order parameter becomes progressively insensitive to variations of the mutation rate. Moreover, for zero backmutation rates, this behavior is this behavior is similar to the one described in \cite{wilke05}, in which the absence of a threshold is argued for. However, for any non-zero backmutation rate, a threshold emerges (see Fig.\ref{figI3}{\bf(b)}).

\section*{Appendix II: Landau Theory of the Transition Matrix Eigenstates}

\begin{figure}[ht!]
  \includegraphics[width=\linewidth]{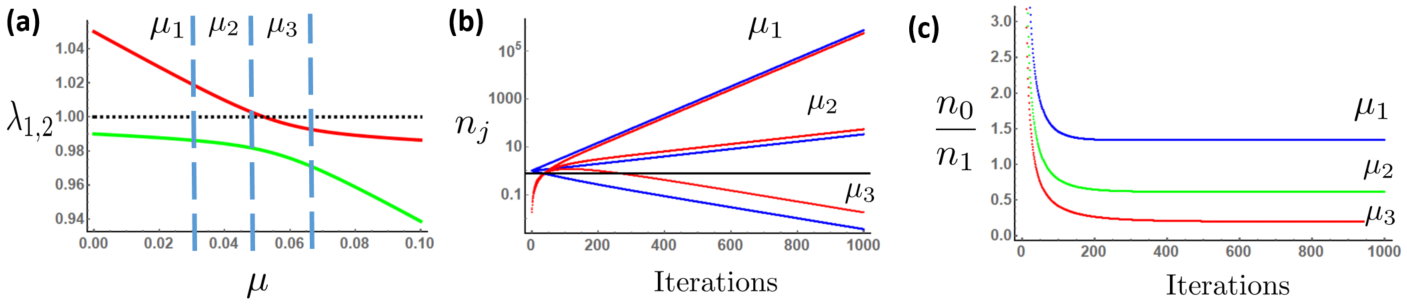}
  \caption{ {\bf(a)} Eigenvalues crossover between $\lambda_1$ and $\lambda_2$ for $f_b = 0.99$. Mutational meltdown happens when $\lambda_1 < 1 $. We consider three mutation rates, $\mu_1 =0.03$, $\mu_1 =0.045$ and $\mu_1 =0.065$ {\bf(b)} Time evolution of the master sequence population $n_0$ (solid blue) and  less fit genotypes $n_1$ (solid red) as a function of time. Interestingly, one observes an inversion just below the mutational meltdown point, due to the existence of an error threshold. {\bf(b)} Ratios for the asymptotic populations for different mutation rates.}
  \label{figI1}
\end{figure}

Even for the shortest viruses, a complete analysis is unfeasible due to the exponential number of configurations. This is not necessarily a problem if we aggregate the genotypes (binary strings) into phenotypes (sets of strings with equivalent functional properties). Indeed, a simple $m+1 \times m+1$ model of the form:

$$
{\cal T}  = \begin{bmatrix}
f(1- \mu) & \mu_B & \dots & \mu_B \\
f\mu /m & 1-\mu_B & \dots & 0 \\
\vdots & \vdots & \ddots & \vdots \\
f\mu /m & 0 & \dots & 1-\mu_B
\end{bmatrix}
$$
allows to obtain the derive a threshold-like behavior. Note that this matrix is not symmetric, and that the rates that connect the master sequence to deleterious or lethal mutations are larger than those for the reverse path. Diagonalising the matrix yields the right eigenvalues (see Fig.\ref{figI1}):

\bqa
\lambda_{max} = \lambda_{1} &=& \frac{1}{2} \left(1-f\mu+f-\mu_B + \sqrt{2 \mu_B (f \mu +f-1)+(f (\mu -1)+1)^2+\mu_B^2}\right) \\
\lambda_{2} &=& \frac{1}{2} \left(1-f\mu+f-\mu_B - \sqrt{2 \mu_B (f \mu +f-1)+(f (\mu -1)+1)^2+\mu_B^2}\right)\\
\lambda_{3:m} &=& 1-\mu_B
\eqa

The dominant eigenvector $ v^{max}$ of the fitness matrix allows to compute the molar fractions of the different genotypes. This vector grows at the fastest rate and rapidly becomes the only relevant one at long times, \ie given the evolution matrix $U(t) = \exp (t {\cal T})$, 

$$
\lim_{t \gg 1} \frac{U(t) v^{max}}{\sum_j U(t) v^{j}} \Longrightarrow 1 (0)  \textrm{ if } \mu < \mu_C  (\mu > \mu_C)
$$
which allows to assume a large and constant viral populations. We define the order parameter $\rho = v^{max}_0 / \sum_i v^{max}_i$, as the ratio between the number of viruses with master sequence genotype against that of all the other sequences. For $f = 1 + s$, one can expand the order parameter $\rho$ to first order in $\mu_B$ to obtain the dependence on temperature:

$$
\rho \approx \frac{\mu_B}{s \ln^2\left(\frac{s}{1+s}\right) \left(T-T_C\right)} + O(\mu^2_B)
$$
which yields a divergent behavior of the susceptibilty at $T_C$, since $\chi = \partial \rho / \partial \mu_B \propto 1/(T-T_C)$ for small backmutation rates. This is in good qualitative agreement with the one expected from a functional maximisation in Landau theory $F(\rho, T, \mu_B)$. To see this, we calculate the ``zero-backmutation" susceptibility for the fitness functional: 

\bqa
\frac{\partial F}{ \partial \rho} = 0 \rightarrow 0 &=& \mu_B - \alpha \rho - \beta  \rho ^3 \\
0 &=& 1 - \alpha \chi - 3 \beta  \rho ^2 \chi\eqa
where we have taken the derivative w.r.t $\mu_B$ in the last line to obtain an expression for the susceptibility. Since $\rho = 0$ above the critical temperature, and $\rho = \sqrt{|\alpha|/\beta}$ below it, we obtain that:

$$
\chi = \frac{1}{\alpha} \left( \frac{1}{2 |\alpha |}\right) \textrm{  for  } \mu > \mu_C \textrm{ }(\mu < \mu_C)
$$

This, incidentally, allows us to derive an analytical form for $\alpha_0$ at the critical temperature:

$$
a_0 = s \ln^2\left(\frac{s}{1+s}\right)
$$

At this point it is important to distinguish the ``Error Catastrophe" from ``Mutational Meltdown" \cite{wagner93,wilke05}. Whereas the former occurs in large populations but is driven by high mutation rates, the latter involves a gradual accumulation of deleterious mutations facilitated by genetic drift. Mutational meltdown entails a gradual decline of the population (\ie the fitness eigenvalue is less than one), whereas the Error Catastrophe is a rapid collapse of genetic information due to a phase transition. Throughout this work, we only consider the mutation dynamics in which fitness remains larger than 1.

\section*{Appendix III: Signatures of Replica Symmetry Breaking}

The concept of Replica Symmetry Breaking (RSB) was introduced to describe the complex energy landscape of spin glasses, and it is a fundamental idea in the study of complex systems and statistical mechanics \cite{mezard87, nishimori11b}. The overlap $q_{SG}$ between two spin configurations $\sigma$ and $\sigma'$ is defined as:

$$
q_{SG} = \frac{1}{L}\sum^L_i \sigma_i \sigma'_i
$$
where $L$ is the number of spins and $\sigma_i \in \{-1,1\}$. In a system exhibiting RSB, the distribution of overlaps (the similarity between different microscopic configurations) is non-trivial. Different instantiations of the quenched disorder will give rise to different overlap distributions, which must be averaged over the replicas, i.e $\bar P(q_{SG}) = \langle P(q_{SG})\rangle_R$.  This means the overlap distribution function over several replicas, $\bar P(q_{SG})$, is not a simple delta function as it would be in a system with replica symmetry, but rather has a broad, almost continuous shape. RSB indicates that the phase space of the system is divided into many metastable states, a hallmark of ergodicity breakdown.\\

In order to apply the methods to study RSB to epistatic fitness landscapes such as the ones that arise from the simplified NK model presented previously, we introduce the overlap between two genotypes $s$ and $s'$ as:

$$
q(s,s') = 1 - d_\textrm{Hamming}(s, s')
$$
where in this case the genomes are bitstrings of length $L$ and $d_\textrm{Hamming}(s, s') = \sum_j (1-\delta_{s_j, s'_j}) / L$ is the Hamming distance. The signature of RSB is the gradual spread of the overlap distribution (see Fig.\ref{fig4} in the main text).\\

In the single-peak case, it is possible to empirically measure the ``mutational susceptibility" due to fluctuations of the conjugate variable. If we call $\langle \rho\rangle$ the average molar fraction: 

\bqa
\chi &=& \frac{\partial \langle \rho \rangle}{\partial \mu_B} = \frac{\partial }{\partial \mu_B}  \left(\frac{1}{\mathcal{Z}} \sum_s p(s) \rho_s \right)   \\
&=&  \frac{1}{T} \left(\frac{1}{\mathcal{Z}} \left(\sum_{s} e^{-(F_{s} - \rho\mu_B)/T}  \rho^2_s\right) -   \frac{1}{\mathcal{Z}^2}\left(\sum_s  e^{-(F_{s} - \rho\mu_B)/T} \rho_s \right)\left(\sum_{s'}e^{-(F_{s'} - \rho\mu_B)/T}\rho_{s'}\right)\right) \\
&=& \frac{1}{T} ( \langle \rho^2\rangle  - \langle \rho\rangle^2)
\eqa
where we have assumed that the population is described by an equilibrium distribution, \ie $p(s)\mu(s\rightarrow s') = p(s')\mu(s '\rightarrow s)$, such that $p(s) = \exp -(F_s - \rho\mu_B)/T$ and $\mathcal{Z} = \sum_s p(s)$.

For spin glasses and rugged lanscapes, magnetisation cannot be used as an order parameter due to lack of long-range order, meaning that due to competing interactions and disorder, $\langle M \rangle = 0$ both in the paramagnetic and the spin-ice phases \cite{nishimori11b}. The alternative is to define the overlap order parameter.

$$
q_{ij} = \frac{1}{L} \sum_k s^{(i)}_k s^{(j)}_k
$$
where $s^{(i)}$ is the genotype associated to the fitness peak $i$. 

In magnetic systems, the conjugate magnitude to this new order parameter is the ``replica field" $r_{ij}$, which is a theoretical construct that allows to write a ``replica Hamiltonian":

$$
H_\textrm{replica} = H_\textrm{magnetic} - \sum_{ij} r_{ij}q_{ij}
$$
which, in paramagnetic phase (\ie the strong mutation weak selection regime), approaches the magnetic Hamiltonian, since the overlaps vanish and  $H_\textrm{magnetic} \gg \sum_{ij} r_{ij}q_{ij}$. Indeed, the magnetisation can be obtained as a limiting case of the overlap order parameter when there is only one energy minimum at $s^{(i)}$:

$$
q_{ii} = \frac{1}{L}\sum_j s^{(i)}_j s^{(i)}_j
$$

The ``spin glass susceptibility", computed as the variance of the overlap distribution variance, measures how the overlap distribution changes due to variations in a replica field (note that the ``replica field" is not a physical quantity, and ultimately the driving forces are magnetic fields and the quenched disorder, which are the counterpart to epistatic intearctions).  This susceptibility captures inter-replica correlation rather than just correlations between spins, as does the magnetic susceptibility. Fluctuations in the overlaps stem from randomness across different instantiations of the epigenetic landscape (replicas). Given the replica-averaged overlap distribution $\bar P(q) = \langle P(q) \rangle_R$, we define:

$$
\chi_{EPI}(T) = \frac{1}{T}\left( \langle q^2 \rangle_{\bar P} -  \langle q \rangle_{\bar P}^2\right)
$$

This ``epistatic susceptibility" will take on a range of values depending on whether we consider intra-phenotype epistasis (that is, we only consider fitness peaks which are clumped into a single phenotype), or whether the overlaps are taken between fitness peaks at arbitrary (Hamming) distance.\\

 It is possible to transition from one regime to the other by imposing a cutoff on the initial overlap ditribution, which sets a upper bound on the fluctuations of the overlap distribution.

$$
\chi_{EPI}(T, q_\textrm{cutoff}) =  \chi_{EPI}(T) \textrm{ calculated for }  \langle P(q) \rangle_R \textrm{ with } q>q_\textrm{cutoff}
$$

Whereas at high temperature the behavior of $\chi_{EPI}$ is qualitatively equivalent to a paramagnetic, disordered phase, at low temperatures, it depends strongly on the ratio between the cutoff $q_\textrm{cutoff}$ and the percentage of the loci that undergo epistatic interactions. For a fixed espistasis probability, at relatively low cutoffs, \ie $q_\textrm{cutoff} \leq 0.7-0.8$ for the parameters in this work, the susceptibility diverges since in the spin-ice phase the variance of the overlaps remains finite (and therefore $\chi_{EPI}$ diverges in the limit $T\rightarrow 0$). This behavior is observed in spin glasses with weak disorder \cite{andreanov10}. On the other hand, at high cutoffs the variance $\langle q^2\rangle_R$ remains small since very similar chains are being compared. In the limit $q_\textrm{cutoff}\rightarrow 1$ only one fitness peak is considered. 

This is reminiscent of  the differentiation into ``linear response susceptibility", which measures responses of spins to small changes in the magnetic field such that the system remains in one energy minimum, and ``equilibrium susceptibility", which measures response over times in which the system has time to accomodate to the lowest energy for a fixed magnetic field and the system becomes effectively insensitive to variations of the external field \cite{parisi23}.

\end{document}